# Smoothed particle hydrodynamics study of friction of the coarse-grained $\alpha$-Al$_2$O$_3$/$\alpha$-Al$_2$O$_3$ and $\alpha$-Fe$_2$O$_3$/$\alpha$-Fe$_2$O$_3$ contacts in behavior of the spring interfacial potential

Le Van Sang[a,*], Akihiko Yano[b], Ai Isohashi[b], Natsuko Sugimura[a,d], Hitoshi Washizu[a,c,**]

[a]Graduate School of Simulation Studies, University of Hyogo - Kobe, Hyogo 650-0047, Japan

[b]Mitsubishi Heavy Industries, Ltd. - 2-1-1 Arai-cho Shinhama, Takasago, Hyogo 676-8686, Japan

[c]Elements Strategy Initiative for Catalysts and Batteries (ESICB), Kyoto University - 1-30 Goryo-Ohara, Nishikyo-ku, Kyoto 615-8245, Japan

[d]Faculty of Engineering, Tokyo City University - 1-28-1 Tamazutsumi, Setagaya, Tokyo 158-8557, Japan

[*]Email: levansang82@gmail.com

[**]Email: h@washizu.org

The paper uses the spring potential to present interaction between the coarse-grained interfacial particles of the $\alpha$-Al$_2$O$_3$/$\alpha$-Al$_2$O$_3$ and $\alpha$-Fe$_2$O$_3$/$\alpha$-Fe$_2$O$_3$ contacts in the sliding friction study of these microscale oxides by smoothed particle hydrodynamics simulations. The spring constants of the potential for the particle systems are converted from those of the atomic oxide systems that are yielded by the second order polynomial fits of the surface potential energy probed by molecular dynamics simulations, and are dependent on the particle coarse-graining. It is found that at microscale the friction properties of the oxides are almost independent of the coarse-graining and are the same in the different sliding directions. Even the hardness contacts friction coefficient shows a decrease with increasing intensity of the normal component of the interfacial interaction, originating from stability of the friction force and growth of the normal force. This result is as an implementation for the previous observations of sliding friction of various materials that showed that a drop of friction coefficient with increasing externally applied normal load has originated from deformation of interfaces or occurrence of debris at contact, indicating an unsteady contact.

**Keywords:** Sliding friction; Metallic oxide; SPH simulation; Coarse grained particle

**1. Introduction**

Nature or kind of interfacial interaction potential has been well-known strongly resulting in tribological properties of materials. There is the relationship between friction coefficients and ionic potentials of various oxides reported by Erdemir [1]. The oxides with very small differences in ionic potential may lead to high friction and the oxides with high ionic potentials or highly screened cations show low shear strength and hence high lubricity [1]. Mechanical interactions between the relative sliding contact surfaces are known as the primary features causing friction between the surfaces. Friction force remarkably reducing with a decrease of the mechanical interactions was verified in the experimental results of Sugihara et al. [2]. Molecular dynamics simulations of Mann et al. also stated that friction coefficient of the α-$Al_2O_3$/α-$Al_2O_3$ contact becomes lower when the interfacial interaction potential is chosen to be weaker [3]. Therefore, understandings or modellings of interfacial interaction play a dominant role in sliding friction studies. Probing this interaction kind has been extensively performed for relative sliding of various contact surfaces. Zhu et al. found the same the anisotropic scenario of the graphene-gold interaction potential energy and friction force in sliding of the graphene flake on the gold substrate [4]. Interaction potential form between the contact oxide surfaces ($SiO_2$ or $Al_2O_3$) has been known inextricably linking to the surface roughness and relating to the stochastic nature of stick-slip events [5]. For coarse-grained (CG) model, the interfacial potential of the graphene sheets reduces with increasing the distance between them and starts reaching a stable one at a 0.335 mm distance [6]. The interfacial potential shows a periodic form along the sliding direction [6].

A few studies have used spring potential to present interaction between interfacial particles in smoothed particle hydrodynamics simulations of sliding friction of the CG centimeter-scale solid rocks [7] and the CG microscale iron [8]. In our views, spring force has not ever been carried out for the similar work of CG metallic oxides. Therefore, this work considers spring force for this aim for the CG α-$Al_2O_3$ and α-$Fe_2O_3$ oxides, which have the corundum lattice structure. Effects of the particle coarse-graining and the modeled interfacial potential on friction of the systems are discussed in detail. The coarse-graining in this work (discussed below) is based upon the coarse-graining of $CuO_2$ oxide mentioned in the work of Iwamoto [9]. Spring constants of CG particles are converted from spring constants of atomic oxide systems that are probed by molecular dynamics simulations and are dependent on size of CG particles. All the simulations in this work are employed in smoothed particle hydrodynamics (SPH) approach.

## 2. Method and calculation

### 2.1. Coarse-grained model

Oxide particle coarse-graining in this study is based upon the work of Iwamoto that lumped each unit cell of $Cu_2O$ oxide into one particle [9]. However, in order to investigate a system of micronsize we slightly modify the coarse-graining of Iwamoto by lumping an atomic region of $n_x \times n_y \times n_z$ (number of the unit cells in the x-, y- and z-directions) oxide unit cells into one particle. Each CG particle is located at the center of mass of the corresponding CG atomic region and has mass of $M_{CG} = n_x n_y n_z \left(12 m_{Al\ or\ Fe} + 18 m_O\right)$ for $Al_2O_3$ or $Fe_2O_3$. This CG method converts an atomic oxide system to a regularly particle lattice system whose unit cell is presented by three vectors $\vec{l}_a$, $\vec{l}_b$ and $\vec{l}_c$ with $\vec{l}_a = n_x a \vec{i}$, $\vec{l}_b = n_x b_1 \vec{i} + n_y b_2 \vec{j}$, $\vec{l}_c = n_z c \vec{k}$, $\left(\vec{l}_a, \vec{l}_b\right) = 120^0$, $\left(\vec{l}_a, \vec{l}_c\right) = 90^0$ and $\left(\vec{l}_b, \vec{l}_c\right) = 90^0$, where $\vec{i}$, $\vec{j}$ and $\vec{k}$ are the unit vectors of the three dimensional Cartesian-coordinate system in the x-, y- and z-directions, respectively; $a = 4.7589$ Å, $b_1 = -2.37945$ Å, $b_2 = 4.121328$ Å and $c = 12.991$ Å for α-$Al_2O_3$ [10]; and $a = 5.038$ Å, $b_1 = -2.519$ Å, $b_2 = 4.363036$ Å and $c = 13.772$ Å for α-$Fe_2O_3$ [11]. A particle system of micronsize can be obtained from expansion of the particle unit cell along the three directions.

### 2.2. SPH approach

In this approach, the time evolutions of the density $\rho_i$, the velocity $v_i^\alpha$ and the internal energy $u_i$ of the $i$th particle are presented by the first order derivative equations of time as follows

$$\frac{d\rho_i}{dt} = \sum_{j=1}^{N} m_j \left(\vec{v}_j - \vec{v}_i\right) \vec{\nabla}_i W\left(\vec{r}_{ij}, h_{ij}\right), \quad (1)$$

$$\frac{dv_i^\alpha}{dt} = \sum_{j=1}^{N} m_j \left(\frac{\sigma_i^{\alpha\beta}}{\rho_i^2} + \frac{\sigma_j^{\alpha\beta}}{\rho_j^2} + \Pi_{ij}\right) \nabla_i^\beta W\left(\vec{r}_{ij}, h_{ij}\right), \quad (2)$$

$$\frac{du_i}{dt} = \frac{1}{2} \sum_{j=1}^{N} m_j \left(\frac{\sigma_i^{\alpha\beta}}{\rho_i^2} + \frac{\sigma_j^{\alpha\beta}}{\rho_j^2} + \Pi_{ij}\right) \left(v_j^\alpha - v_i^\alpha\right) \nabla_i^\beta W\left(\vec{r}_{ij}, h_{ij}\right), \quad (3)$$

where $\alpha, \beta \equiv x, y, z$; $m$ is mass of the particle; $\vec{r}_{ij} = \vec{r}_i - \vec{r}_j$ is relative position vector between the particles $i$ and $j$; $h_{ij}$, $\sigma^{\alpha\beta}$ and $\Pi_{ij}$ are smoothed length, stress tensor and artificial viscosity function, respectively, which are considered similarly to those in our previous work [8]; and $W\left(\vec{r}_{ij}, h_{ij}\right) = 1/(\pi^{3/2} h_{ij}^3)\left(5/2 - r_{ij}^2/h_{ij}^2\right)\exp(-r_{ij}^2/h_{ij}^2)$ is a Gauss kernel function. In order to compensate energy dissipation caused by friction during the sliding, a dissipation force is afforced on each particle of the system as follows

$$F_{dis,i} = \begin{cases} -m_i \gamma_{dis} \left(v_i^x - V_{dis}\right) & \text{the x-direction} \\ -m_i \gamma_{dis} v_i^y & \text{the y-direction} \\ -m_i \gamma_{dis} v_i^z & \text{the z-direction,} \end{cases} \quad (4)$$

where $\gamma_{dis}$ is a parameter of the model and $V_{dis} = 0$ for particles of the substrate and $V_{dis} = V$, which is a constant sliding velocity of the slider, for particles of the slider. The Prandtl-Tomlinson model is utilized by adding a spring force on each particle of the slider as follows

$$F_{spr,i} = \begin{cases} K(x_{0,i} + Vt - x_i) & \text{the x-direction} \\ K(y_{0,i} - y_i) & \text{the y-direction} \\ K(z_{0,i} - z_i) & \text{the z-direction,} \end{cases} \quad (5)$$

where $K$ is a spring constant, $t$ is sliding time, $x_{0,i}$, $y_{0,i}$ and $z_{0,i}$ are the equilibrium/initial coordinates of the $i$th particle in the x-, y- and z-directions, respectively. Interaction between the slider and the substrate is presented by interaction between particles of the two particle layers, one of the slider and the other of the substrate, in the contact. Two particles, one of each layer, interact with each other by a spring force as follows

$$\vec{F}_{int,ij} = \begin{cases} -K_\alpha (r - h_{ij}) \dfrac{\vec{r}_{ij}}{r} & 0 < r < h_{ij} \\ 0 & r > h_{ij}, \end{cases} \quad (6)$$

where $K_\alpha$ is a spring constant. The friction force $F_{fri}$, the normal force $F_{nor}$ and the friction coefficient $\mu_{cof}$ are defined as follows

$$F_{fri} = \sum_{i=1}^{N_f} \left( F_{spr,i}^x + F_{int,ij}^x \right), \quad (7)$$

$$F_{nor} = \sum_{i=1}^{N_f} \left( F_{spr,i}^z + F_{int,ij}^z \right), \quad (8)$$

$$\mu_{cof} = \frac{F_{fri}}{F_{nor}}, \quad (9)$$

where $N_f$ is the number of the particles (the friction particles) of the contact layer of the slider, $F^x$ and $F^z$ are the force components in the x- and z-directions, respectively.

*2.3. Spring constant of particle*

The spring constants $K_\alpha$ are converted from spring constants of surface potentials of the atomic oxide systems probed by molecular dynamics (MD) simulations. Fig. 1 presents the Al$_2$O$_3$ atomic system (similarity for Fe$_2$O$_3$) consisting of the two objects used in the MD simulations by LAMMPS [12]. The lower object is fixedly held during all the MD simulations while the upper object is regularly shifted a 0.01 Å step along the z-direction and a 0.02 Å step along the x- or y-direction. The Coulomb + Lennard-Jones interaction potential $U(r_{ij}) = Cq_iq_j / (\varepsilon_{cl} r_{ij}) + 4\varepsilon_{lj}(\sigma_{ij}^{12}/r_{ij}^{12} - \sigma_{ij}^6/r_{ij}^6)$, where $C$ is the energy conversion constant (see in LAMMPS), $\varepsilon_{cl} = 1$ is a dielectric constant, $q_i$ and $q_j$ are charges of the atoms $i$ and $j$, $\varepsilon_{lj}$ and $\sigma_{ij}$ are parameters, is used to present interaction between atoms of the oxides. The parameters $q_i$, $q_j$, $\varepsilon_{lj}$ and $\sigma_{ij}$ for Al$_2$O$_3$ and Fe$_2$O$_3$ are taken from the works of Hu et al. [13], and Kerisit [14] and Olsen et al. [15], respectively. The Lorentz-Berthelot combining rules of the Lennard-Jones potential is applied for $\varepsilon_{lj}$ and $\sigma_{ij}$. Fig. 2 shows the surface potential energy of the oxides in the x-, y- and z-directions. We first monitor the potential energy in the z-direction to find its (first) minimum position, 1.67 Å for Al$_2$O$_3$ or 1.34 Å for Fe$_2$O$_3$. Distance in the z-direction between the two objects of each oxide (Fig. 1) is then chosen to be equally to its minimum position value during shifting the upper object along the x- or y-direction. We obtain the spring constants of the atomic oxide systems in the x- and y-directions from the second order polynomial fits of these potential energy curves (Fig. 2), for example, the spring constants of $k_x^{atom} = 0.5055$ eV/Å$^2$ and $k_y^{atom} = 1.4943$ eV/Å$^2$ are obtained for Al$_2$O$_3$ in the x- and y-directions, respectively. Error of each spring constant is less than 1 % of its value. In order to convert these spring constants to the spring constants of particles ($K_\alpha$), we use a picture of serial connection of the springs in the CG directions. This means that for a given coarse-graining in which a particle is yielded from a group of $n_x$ and $n_y$ unit cells, the relationship of the spring constants between an atomic oxide system and a corresponding CG particle one is presented as follows

$$\frac{1}{K_x} = \frac{1}{k_x^{atom,1}} + \frac{1}{k_x^{atom,2}} + ... + \frac{1}{k_x^{atom,n_x}} = \frac{n_x}{k_x^{atom}}, \quad (10)$$

$$\frac{1}{K_y} = \frac{1}{k_y^{atom,1}} + \frac{1}{k_y^{atom,2}} + ... + \frac{1}{k_y^{atom,n_y}} = \frac{n_y}{k_y^{atom}}. \quad (11)$$

It should be noted that the spring constants of particle are dependent on its size ($n_x$ and $n_y$). This relationship is not considered in the z-direction because atoms lumped into one particle must be belonged to the only object (the slider or the substrate). The spring constant of particle in the z-direction ($K_z$) is adjusted over a range of value until simulation data well approximate with those of friction of the oxides reported in previous studies.

*2.4. Simulation parameters and systems*

Each of the microscale atomic oxide systems, $Al_2O_3$ or $Fe_2O_3$, is coarse-grained into three different particle systems related to difference of the particle size, $n_x$, $n_y$ and $n_z$ (see in Table 1). Quantities of particle such as the mass ($M_{CG}$) and the spring constants ($K_x$ and $K_y$) are then estimated and listed in Table 1. Each of the simulated particle systems consists of a slider and a substrate whose sizes, $L_x$, $L_y$ and $L_z$, in the directions are listed in Table 2.

The number of the unit cells of each CG particle system expanded in each direction to reach its size is listed in Table 3. The particle number of each system is then estimated (Table 3). The spring constant $K$ is taken from the spring constant of the $Al_2O_3$ or $Fe_2O_3$ oxide-functionalized cantilever mentioned in the studies of Gan et al. [16] and Lower et al. [17] (Table 1). Fig. 3 presents the $Al_2O_3$ particle system used in the SPH simulations to detect its sliding friction in the x- and y-directions. The simulated $Fe_2O_3$ particle systems are designed similarly to the $Al_2O_3$ particle system. The initial distance along the z-direction between the slider and the substrate of each particle system is set equally to its $\left|\vec{l}_c\right| = n_z c$. The simulation parameters are set as follows: $\gamma_{dis} = 10^7$ 1/s, density of $\rho_{Al_2O_3} = 3.97$ and $\rho_{Fe_2O_3} = 5.26$ g/cm$^3$, shear modulus of $\mu_{Al_2O_3} = 163.4$ [18] and $\mu_{Fe_2O_3} = 54$ GPa [19], a time step of $\Delta t = 100$ ps, a sliding velocity of $V = 100$ m/s and an external pressure of 3 MPa in all the SPH simulations. The SPH simulation code is modified from the FDPS open source developed by Iwasawa et al. [20].

## 3. Results and discussion

Fig. 4 presents the friction force of the A3000 and F3000 systems in the x- and y-sliding directions with the different ratios of $K_z / K_x$ ranged from 0.01 to 0.12. Each curve shows good periodicity in sliding distance indicating the fact that the contact surfaces have been steady maintained during the sliding. Similarly, stick-slip motion has clearly been observed in relative sliding of hardness contact surfaces [21]. In this work, the systems are steady held by the chosen high value of the dissipation parameter ($\gamma_{dis}$). The hardness contacts make small oscillations of the interfacial particles that result in the small $F_{spr}$ spring forces exerting on the friction particles. In addition, the $\vec{F}_{int}$ spring force in the x-direction very slightly varies at a given value of $K_x$. Consequently, the friction force has no change with varying the ratios of $K_z / K_x$ as seen Fig. 4. It is also founded that the average friction force is not dependent on the sliding directions, being equally to 0.24 µN for the A3000 system and 0.28 µN for the F3000 system in the x-direction or the y-direction. These results match closely to those mentioned in experimental and simulation reports for diamond surfaces. The average friction forces of the (100) and (111) diamond surfaces are found to be similarly and independent of load up to 0.1 µN in the ultrahigh vacuum force microscope measurements of a clean diamond tip sliding on these surfaces along the various directions [22]. Gao et al. used the nanowire diamond AFM tips of different radii to experimentally probe friction force of (001) and (111) diamond surfaces at a zero load and found that with the tip of a certain radius the measured values of friction force of the five different surface-orientation/sliding-direction combinations are near or close the same [23]. Their MD simulations also confirmed that when the error bars are considered, there are no statistically significant differences in friction force between either the surface or for any sliding direction at low pressures; however, with external pressures greater than 35 GPa the average friction force when sliding in the [$\bar{1}10$] direction on the (001) diamond surface is always smaller than when sliding in the [110] direction [23]. As an outlook, average friction force between hardness contact surfaces is independent of sliding direction in behavior of small normal component of interfacial interaction such as $K_z$ in this work and applied load or external pressure in the previous studies [22,23]. Similarly, the periodic curves of the friction force are found for the other systems, two of them (A1000 and F1500) are presented in Fig. 5. The average friction force increases as size of CG particle decreases, as seen from comparison between the curves (and also curves of the other CG systems, not shown) of Figs. 4 and 5. Since the increase of the average friction force is not proportional to $K$ or the increase of $K_x$, both the spring forces, one of the Prandtl-Tomlinson model and the other of the interfacial interaction, have significant contributions to friction force. This indicates that oscillation of particles along sliding direction plays a significant contribution in estimation of friction force. With smaller size of GC particle, the friction force curves show a larger reflection among them (Figs. 4 and 5).

The normal force shows good periodicity in sliding distance as seen in Fig. 6. The average normal force varies proportionally to change of the $K_z/K_x$ ratio in each of the four behaviors. This indicates that the spring interfacial force (Eq. (6)) mainly contributes to normal component of force exerting on the friction particles. Therefore, potential kind chosen to present interaction between interfacial CG particles plays an important role in simulation studies of sliding friction, as also mentioned for an atomic system [3]. At a given $K_z/K_x$ ratio there are the similar values of the average normal force in the x- and y-sliding directions for the system A3000 or F3000 indicating the fact that the average normal force is almost independent of sliding direction. By investigating relative sliding of the (100)-(2×1) reconstructed diamond surfaces at low applied load, Perry et al. found good periodicity of both friction and normal forces in sliding distance and the average normal forces of 0.20 nN/atom and 0.19 nN/atom for $[0\bar{1}1]$ and $[011]$ sliding directions, respectively [21]. Briefly, under condition of low applied load or small modeled normal interaction, friction and normal forces of hardness contacts of atomic system or CG particle system show very little varying along different sliding directions. However, it should be noticed that because of lattice structure the lower peaks additionally appear to interleave the higher peaks in the friction force curves as seen in this work (Fig. 4) and the work of Pang et al. [24] that simulated the atomic-scale friction of a typical two-dimensional single-layer molybdenum disulfide. This work also states that at micron-scale there is a little difference of the average friction and normal forces between the CG $Al_2O_3$ and $Fe_2O_3$ oxides. Good periodicity of both the friction and normal forces indicates a steady state of the systems in the sliding and normal directions during the sliding.

As a result of changes of the friction and normal forces, the friction coefficient also shows a periodic form in sliding distance (Fig. 7). For a given system and a given sliding direction, shape of the curves is the same regarding of the considered $K_z/K_x$ ratio. The oscillation amplitude and average value of the friction coefficient, considered as the friction coefficient of the system in the following discussions, rapidly reduce with increasing the $K_z/K_x$ ratio up to 0.05 and become stable at the remaindering $K_z/K_x$ ratios (Figs. 7 and 8). This state is originated from the change of the normal force. At the contact, the substrate surface always pushes the friction particle layer far away from it because the positive normal force is observed during the sliding (Fig. 6). Therefore, an increase of the $K_z$ can be considered as approximation of an increase of externally applied normal load as long as contact surface of a slider is pushed stronger as external load is applied larger. In this approximation, the results of this work are in accordance with those reported in the previous studies for external load (or pressure) dependence of friction coefficient. In a studied interval of load, friction coefficient of various materials is founded to markedly drop at lower loads and stabilize at higher loads [25,26]. Deformation of interfaces or occurrence of debris at contact has been considered as features for explaining this dependence [27]. However, this work finds that the dependence still appears in sliding friction of hardness contacts with increasing intensity of normal component of the interfacial spring interaction. This result is as an additional understanding for the load-friction coefficient dependence of contacts in various states. It is clearly seen in Fig. 8 that at micron-scale the friction coefficient of the systems almost does not depend on both the sliding direction and the CG particle size, only a small reflection occurring in the CG systems of the α-$Al_2O_3$ oxide. The friction coefficient of reducing from 0.24 to 0.09 in the stable region ($K_z / K_x \geq 0.05$) observed for all the CG α-$Al_2O_3$ systems is in accordance with the experimentally unlubricated sliding friction coefficient of the alumina/alumina or sapphire/sapphire contact such as 0.07 for the smoothed contact [28], $0.08 \pm 0.02$ for the rough contact of $R_a = 2$ nm [29] and about from 0.15 to 0.19 for the rough contact of $R_a = 20$ nm [30], or the MD simulation value of 0.12 for the pure α-$Al_2O_3$ (0001)/α-$Al_2O_3$(0001) contact held at 300 K [31]. The friction coefficient of the CG α-$Fe_2O_3$ systems in the stable region reduces from 0.21 to 0.09 being not agreeably with the steady value of 0.60 [32] reported for the $Fe_2O_3$ ball/$Fe_2O_3$ flat pairs. An agreeable value of the friction coefficient with the reported value can be seen around the $K_z/K_x$ ratio of 0.02. For the two CG $Al_2O_3$ and $Fe_2O_3$ oxides, the friction coefficient reduces with increasing intensity of the interfacial interaction in normal direction. This observation is different from the MD simulation observation of Mann et al. that showed that friction coefficient of the α-$Al_2O_3$/α-$Al_2O_3$ contact is lower with using a weaker modeled interfacial potential [3].

## 4. Conclusions

The paper uses the spring potential to present interaction between the interfacial particles of the α-Al2O3/α-Al2O3 and α-Fe2O3/α-Fe2O3 contacts in the sliding friction study of the CG micron-scale $Al_2O_3$ and $Fe_2O_3$ oxides by the smoothed particle hydrodynamics simulations. Some conclusions are pointed out as follows. The coarse-graining does not result in the observed friction properties of the oxides. At micron-scale, the monitored friction quantities are not dependent on the different sliding directions. With the hardness contacts that show the steady states during the sliding, the friction coefficient still reduces with increasing intensity of normal component of the interfacial force. This result is as an implementation for the previous observations of sliding friction of various materials that showed that a drop of friction coefficient with increasing externally applied normal load has originated from deformation of interfaces or occurrence of debris at contacts, indicating unsteady contacts. As compared to the results of the previous studies, the friction coefficient of the α-$Al_2O_3$ oxide should be taken in its stable region ($K_z/K_x \geq 0.05$), while that of the α-$Fe_2O_3$ oxide should be taken prior to its stable region.

**Table captions**

**Table 1**

Properties of particle: $n_x$, $n_y$ and $n_z$ are the cell unit numbers of the atomic oxide region in the x-, y- and z-directions coarse-grained into one particle; $M_{CG}$ ( unit of $10^{-6}$ μg) is mass of particle; and $K$, $K_x$ and $K_y$ (unit of $10^{-3}$ N/m) are the spring constants of particle. Noting for symbol of System, "A" (as in A1000) typical for the CG system of $Al_2O_3$ with $n_x = 1000$ or $n_y = 1000$, similarly, "F" (as in F1000) typical for the CG system of $Fe_2O_3$.

**Table 2**

Size of the CG oxide particle systems (or atomic systems) in the directions used in the SPH simulation (unit of μm).

**Table 3**

Characters of the simulated particle systems: $N_x$, $N_y$ and $N_z$ are the unit cell numbers of the particle systems expanded in the directions; and $N_{CG}$ is the particle number.

**Figure captions**

**Fig. 1.** The initially $Al_2O_3$ atomic system used to probe the surface potential in MD simulations. The upper object of 6000 atoms and the lower object of 384000 atoms. Red for O and yellow for Al. The atomic oxide systems ($Al_2O_3$ and $Fe_2O_3$) are created from the CIF files taken from American Mineralogist Crystal Structure Database.

**Fig. 2.** The surface potential energy of the atomic oxide systems in the directions probed by MD simulations. Red lines are their second order polynomial fits.

**Fig. 3.** The A1000 particle system used in SPH simulations to detect its friction in the sliding along the x- and y-directions.

**Fig. 4.** The friction force dependent on the sliding distance for the A3000 and F3000 systems. Color lines from left to right for $K_z/K_x = 0.01-0.12$ with increment of 0.01.

**Fig. 5.** The friction force dependent on the sliding distance for the A1000 and F1500 systems. Color lines are the same those noted in Fig. 4.

**Fig. 6.** The normal force dependent on the sliding distance for the A3000 and F3000 systems. Color lines are the same those noted in Fig. 4.

**Fig. 7.** The force coefficient dependent on the sliding distance for the A3000 and F3000 systems. Color lines are the same those noted in Fig. 4.

**Fig. 8.** The friction coefficient dependent on the $K_z/K_x$ ratio for all the simulated particle systems.

**Table 1**

| System | $n_x$ | $n_y$ | $n_z$ | $M_{CG}$ | $K$ | $K_x$ | $K_y$ |
| --- | --- | --- | --- | --- | --- | --- | --- |
| A1000 | 1000 | 1000 | 550 | 0.55881 | 60[16] | 8.08808 | 23.91018 |
| A1500 | 1500 | 1500 | 550 | 1.25732 | 60[16] | 5.39205 | 15.94012 |
| A3000 | 3000 | 3000 | 550 | 5.02929 | 60[16] | 2.69602 | 7.97006 |
| F1000 | 1000 | 1000 | 550 | 0.87535 | 70[17] | 13.40801 | 4.42967 |
| F1500 | 1500 | 1500 | 550 | 1.96953 | 70[17] | 8.93867 | 2.95311 |
| F3000 | 3000 | 3000 | 550 | 7.87815 | 70[17] | 4.46933 | 1.47655 |

**Table 2**

| System | Object | $L_x$ | $L_y$ | $L_z$ |
|---|---|---|---|---|
| CG Al$_2$O$_3$ | Slider/Substrate | 41.759/106.004 | 24.109/61.201 | 6.431/6.431 |
| CG Fe$_2$O$_3$ | Slider/Substrate | 44.207/112.221 | 25.523/64.791 | 6.817/6.817 |

**Table 3**

| System | Object | $N_x$ | $N_y$ | $N_z$ | $N_{CG}$ |
|---|---|---|---|---|---|
| A1000 | Slider/Substrate | 60/150 | 60/150 | 10/10 | 36000/225000 |
| A1500 | Slider/Substrate | 40/100 | 40/100 | 10/10 | 16000/100000 |
| A3000 | Slider/Substrate | 20/50 | 20/50 | 10/10 | 4000/25000 |
| F1000 | Slider/Substrate | 60/150 | 60/150 | 10/10 | 36000/225000 |
| F1500 | Slider/Substrate | 40/100 | 40/100 | 10/10 | 16000/100000 |
| F3000 | Slider/Substrate | 20/50 | 20/50 | 10/10 | 4000/25000 |

**Fig. 1**

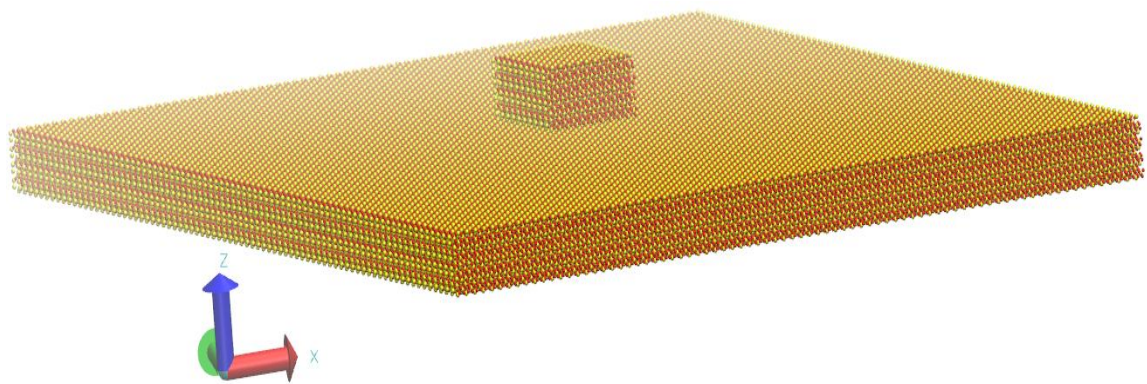

Fig. 2

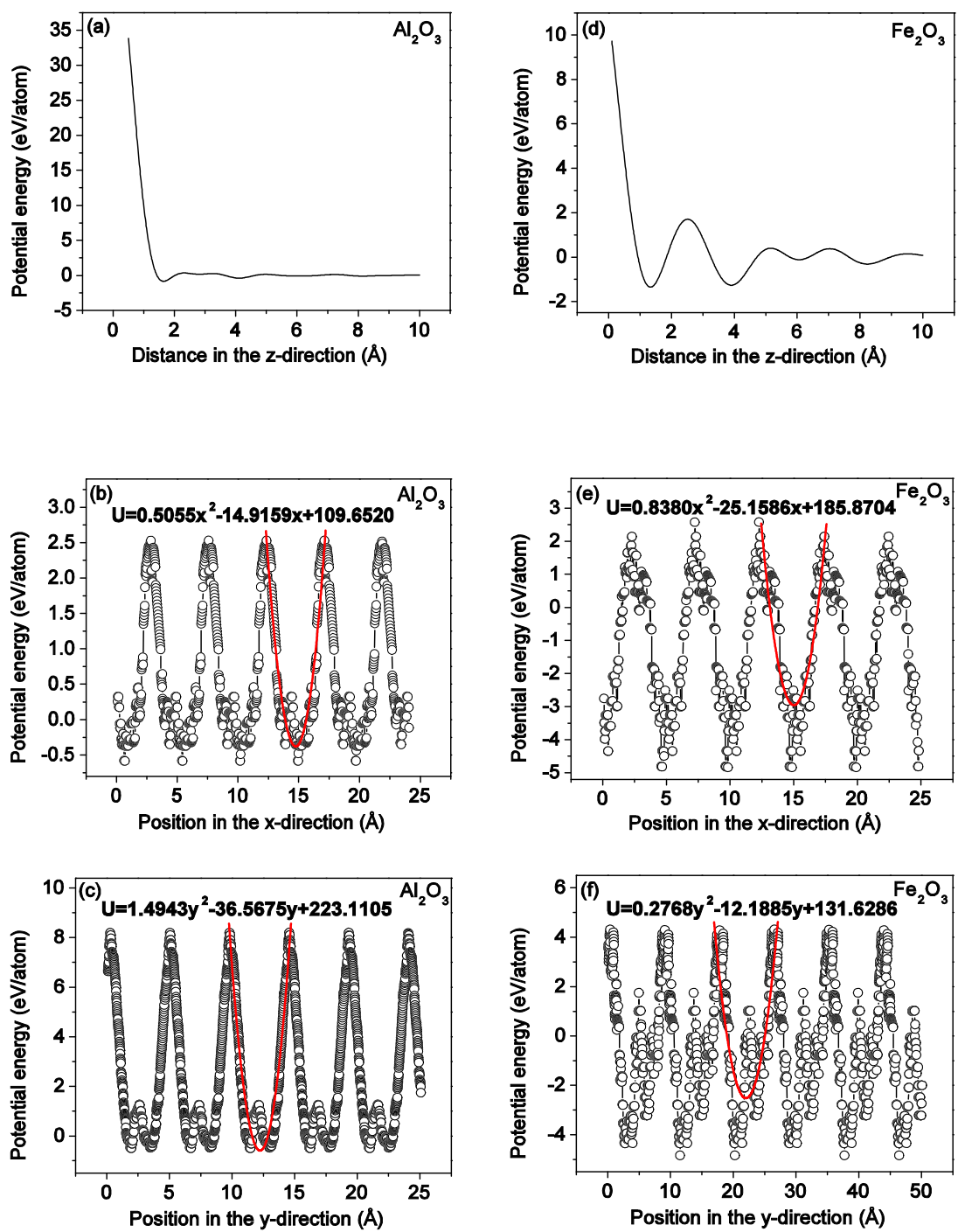

**Fig. 3**

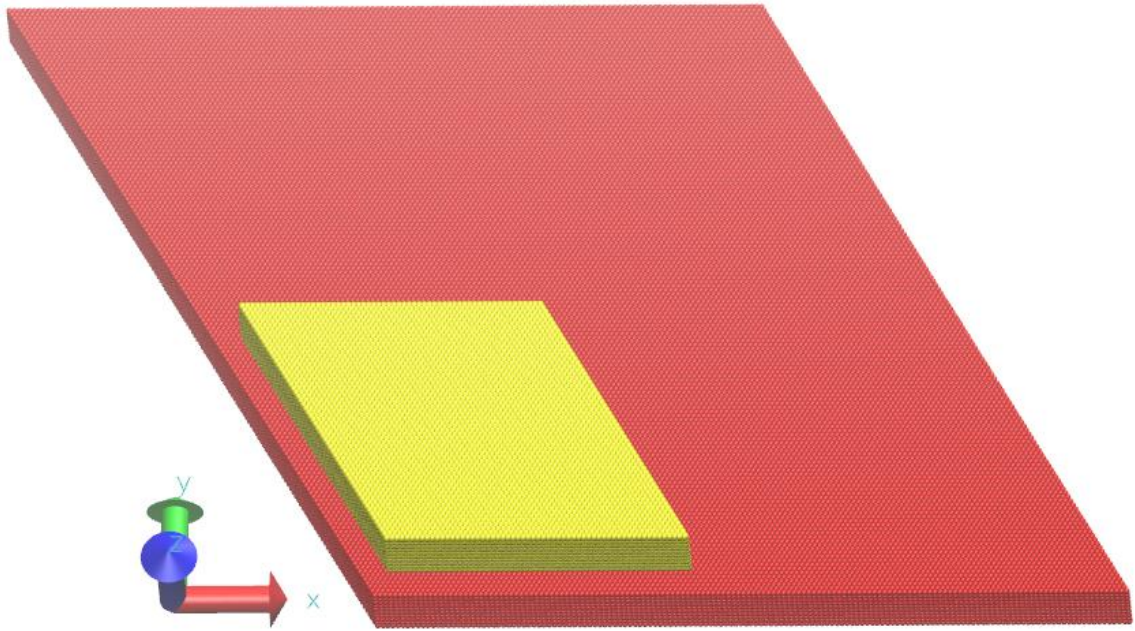

**Fig. 4**

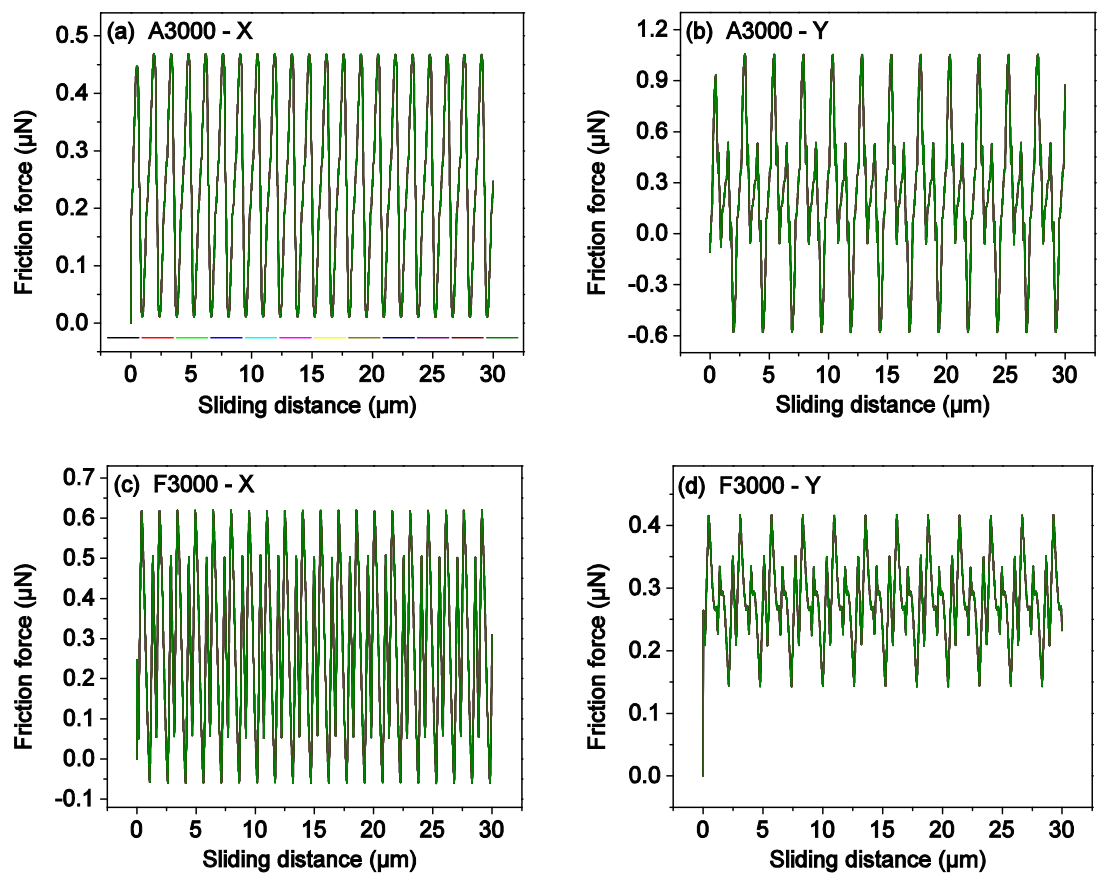

Fig. 5

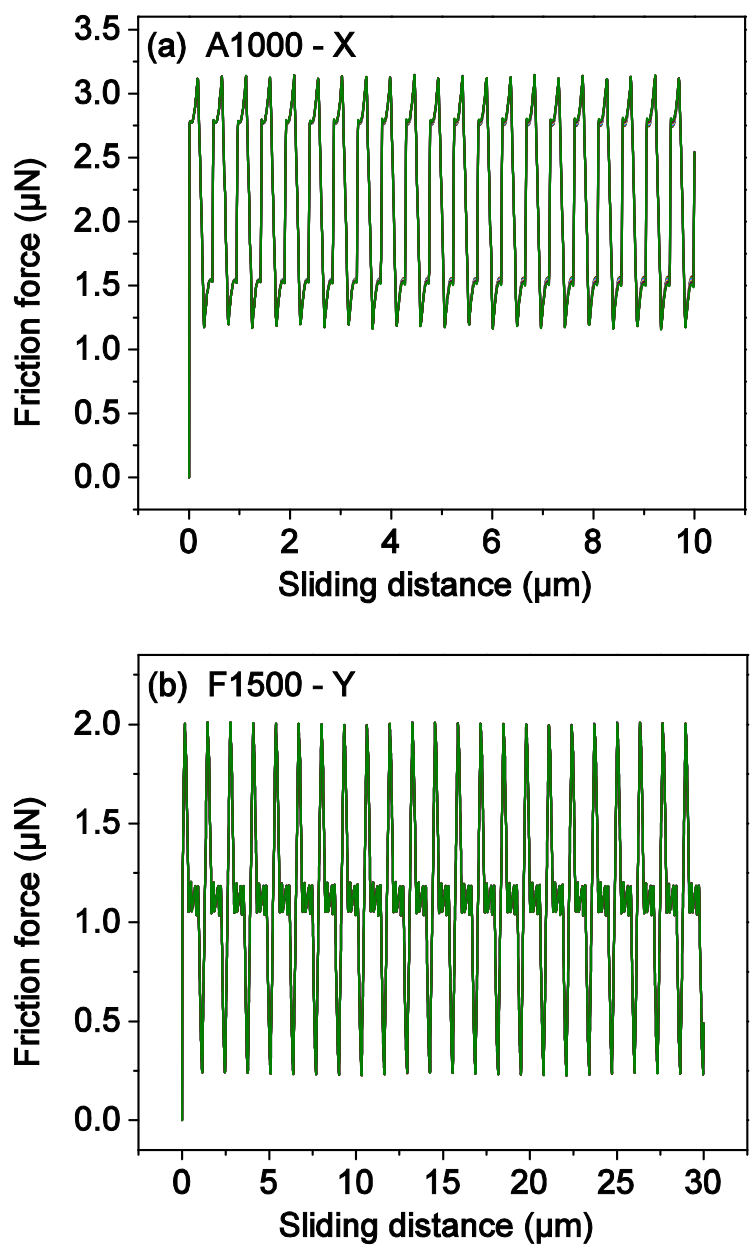

**Fig. 6**

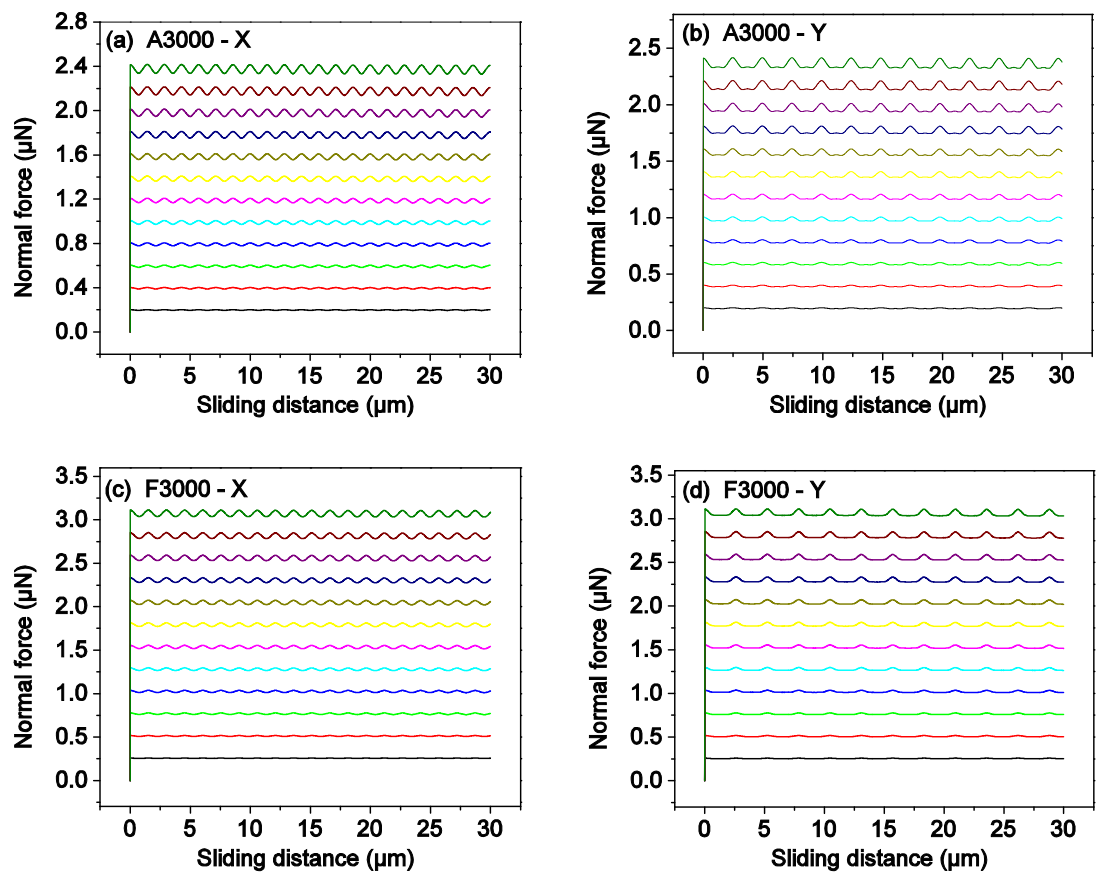

**Fig. 7**

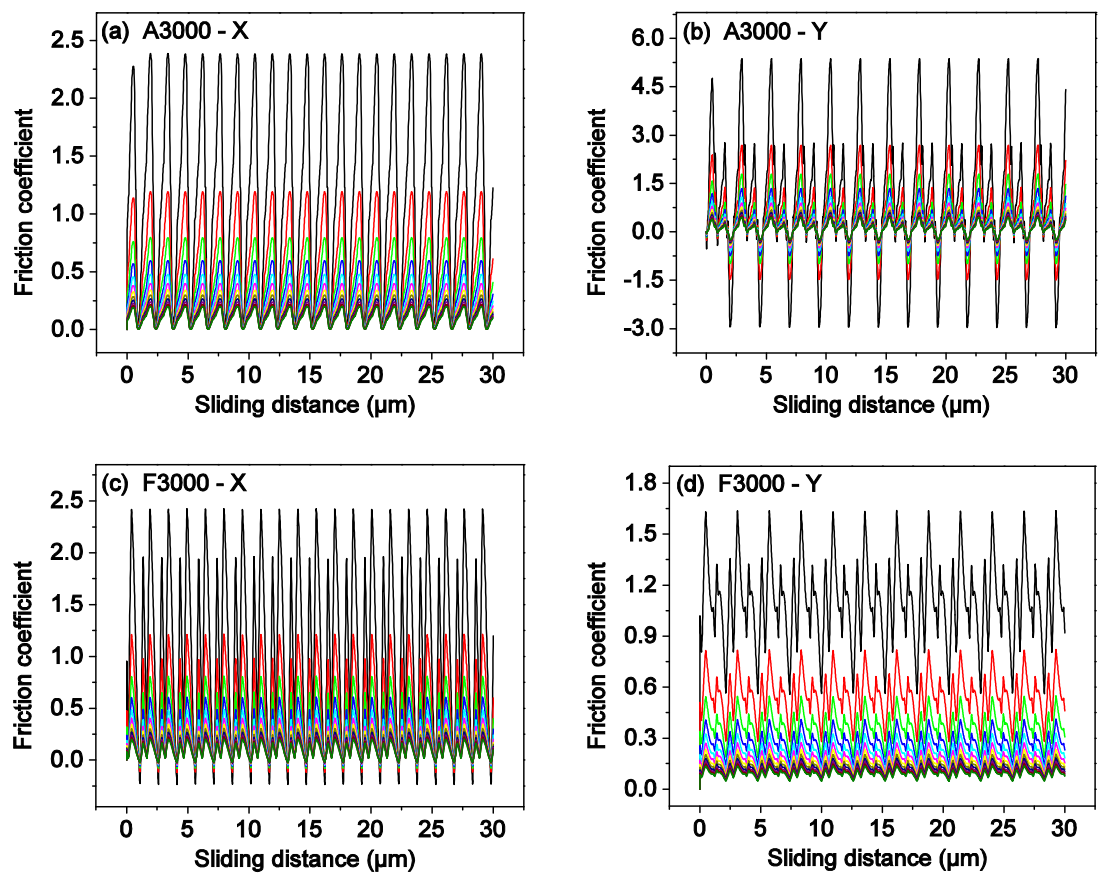

Fig. 8

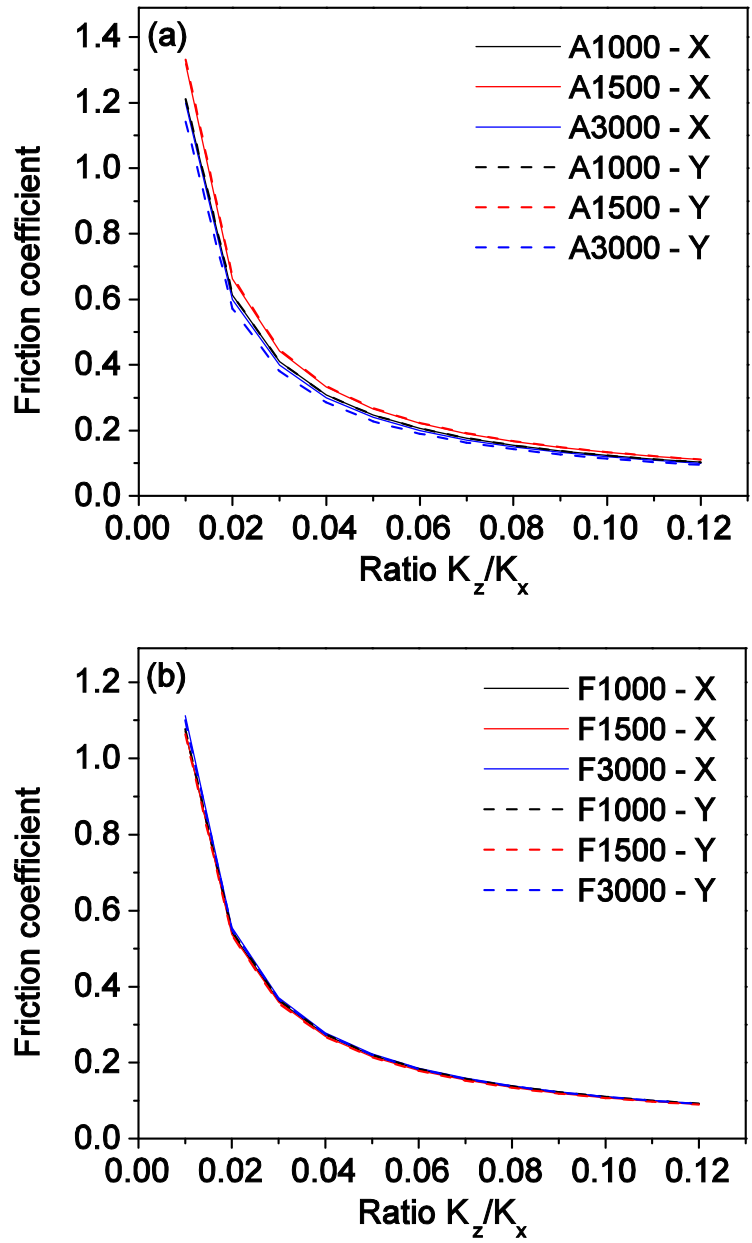